\documentclass[11pt]{article}             
\usepackage{osid}                         %
\usepackage{cite}                         %
\usepackage{amsmath}

\newcommand{\tr}{\mathop{\mathrm{tr}}\nolimits}
\renewcommand{\d}{d}
\newcommand{\e}{e}
\newcommand{\rint}{\int}

\title{Derivation of Master Equations in the Presence of Initial 
Correlations with Reservoir: Projection Method Revisited}
\author{Kazuya Yuasa\\[0.7truemm]
{\footnotesize
\textsl{Research Center for Information Security, 
National Institute of Advanced Industrial Science and Technology (AIST), 1-18-13 Sotokanda, Chiyoda-ku, Tokyo 101-0021, Japan}\\[1truemm]
\textsl{E-mail: kazuya.yuasa@aist.go.jp}\\
}}
\date{April 14, 2006}

\begin{document}
\maketitle
\begin{abstract}
We discuss the derivation of master equations in the presence of 
initial correlations with the reservoir.
In van Hove's limit, the total system behaves as if it started 
from a factorized initial condition.
A proper choice of Nakajima--Zwanzig's projection operator is 
crucial and the reservoir should be endowed with the mixing 
property.
\end{abstract}

\section{Introduction}
The derivation of master equations has been a central issue for 
decades in the field of quantum physics, both from a practical and 
fundamental point of view.
The interests include dissipative dynamics in quantum optics 
\cite{ref:MilburnTextbook,ref:QuantumNoise3rd}, understanding 
decoherence \cite{ref:JoosTextbook}, the emergence of 
irreversibility \cite{ref:QuantumNoise3rd,ref:KuboTextbook}, and 
the influence of dissipation/decoherence on quantum information.

It is not a trivial problem how to describe the dissipative and  
irreversible dynamics of quantum systems, since the Schr\"odinger 
equation applies to closed systems and describes unitary 
evolutions.
In the standard approach to this issue, one puts the system in a 
large reservoir to allow energy exchange and dissipation.
Then, one applies the Schr\"odinger equation to the total 
(closed) system, system S+reservoir B, and extracts the dynamics 
of system S by tracing out the reservoir degrees of freedom,
\begin{equation}
\rho_\text{S}(t)=\tr_\text{B}\rho(t),
\label{eqn:PartialTrace}
\end{equation}
which exhibits a non-unitary evolution and is described by a 
master equation.
Here, $\rho(t)$ and $\rho_\text{S}(t)$ are the density operators 
of the total system and system S, respectively.

One of the targets of this article is to discuss a hypothesis 
that is taken for granted in most of such approaches: factorized 
(uncorrelated) initial conditions
\begin{equation}
\rho_0=\rho_\text{S}\otimes\rho_\text{B}
\label{eqn:FactorizedInitialState}
\end{equation}
are usually assumed in the derivation of master equations without 
any reasoning about the validity and the necessity of this 
hypothesis \cite{ref:MilburnTextbook,ref:QuantumNoise3rd,ref:JoosTextbook,ref:KuboTextbook}.
We will discuss in the following a derivation of master equations 
for generic (correlated) initial states.
The reconsideration of the factorized initial condition 
(\ref{eqn:FactorizedInitialState}) raises another question in the 
application of Nakajima--Zwanzig's projection operator 
$\mathcal{P}$ \cite{ref:QuantumNoise3rd,ref:KuboTextbook,ref:Nakajima,ref:HaakeGeneralizedMasterEq}, defined by the mapping
\begin{equation}
\rho(t)\longrightarrow\mathcal{P}\rho(t)
=\tr_\text{B}\{\rho(t)\}\otimes\Omega_\text{B}.
\label{eqn:P}
\end{equation}
Which state of reservoir B should be taken as the reference state 
$\Omega_\text{B}$ of the projector $\mathcal{P}$?
We address these points and show the derivation of master 
equations in the presence of initial correlations.
We clarify that the choice of $\mathcal{P}$ is crucial and the 
reference state $\Omega_\text{B}$ should be endowed with the 
mixing property \cite{ref:Factorize}.

\section{Framework}
Let the total system consist of a small system S and a large 
reservoir B\@.
The Hamiltonian of the total system reads
\begin{equation}
H=H_0+\lambda H_\text{SB}
=H_\text{S}+H_\text{B}+\lambda H_\text{SB},
\end{equation}
where $H_\text{S}$, $H_\text{B}$, and $H_\text{SB}$ are the 
Hamiltonians of system S, reservoir B, and the interaction 
between them, respectively, and $\lambda$ is the coupling constant.
The corresponding Liouvillians are defined as the commutators 
with the Hamiltonians,
\begin{equation}
\mathcal{L}=-i[H,{}\cdot{}]
=\mathcal{L}_0+\lambda\mathcal{L}_\text{SB}
=\mathcal{L}_\text{S}+\mathcal{L}_\text{B}
+\lambda\mathcal{L}_\text{SB}.
\end{equation}
Clearly,
\begin{equation}
[\mathcal{L}_\text{S},\mathcal{L}_\text{B}]=0.
\end{equation}
We assume that the system Hamiltonian $H_\mathrm{S}$ admits a pure
point spectrum, and the system Liouvillian $\mathcal{L}_\mathrm{S}$
is resolved in terms of its eigenprojections $\tilde{Q}_m$,
\begin{equation}
\mathcal{L}_\mathrm{S}=-i\sum_m \omega_m\tilde{Q}_m,\qquad
\sum_m\tilde{Q}_m=1,\qquad
\tilde{Q}_m\tilde{Q}_{n}=\delta_{mn}\tilde{Q}_m.
\label{eqn:eigenA9bis}
\end{equation}

In this article, we discuss the derivation of master equations 
via Nakajima--Zwanzig's projection method, to implement the trace 
over the reservoir (\ref{eqn:PartialTrace}).
We define two projection operators: $\mathcal{P}$ defined by 
(\ref{eqn:P}) and
\begin{equation}
\mathcal{Q}=1-\mathcal{P}.
\end{equation}
They are projection operators since they satisfy
\begin{equation}
\mathcal{P}+\mathcal{Q}=1,\quad
\mathcal{P}^2=\mathcal{P},\quad
\mathcal{Q}^2=\mathcal{Q},\quad
\mathcal{P}\mathcal{Q}=\mathcal{Q}\mathcal{P}=0,
\end{equation}
assuming the normalization of the reference state 
$\tr_\text{B}\Omega_\text{B}=1$.
One of the main issues of the present article is to discuss the 
choice of the reference state $\Omega_\text{B}$.
We do not know at this moment which state should be taken as the 
reference state $\Omega_\text{B}$.
We only assume its stationarity here, 
$\mathcal{L}_\text{B}\Omega_\text{B}=0$, and will specify other 
features later.

Notice that
\begin{equation}
[\mathcal{P},\mathcal{L}_\text{S}]=0,\quad
[\mathcal{Q},\mathcal{L}_\text{S}]=0,\quad
\mathcal{P}\mathcal{L}_\text{B}
=\mathcal{L}_\text{B}\mathcal{P}=0,\quad
\mathcal{Q}\mathcal{L}_\text{B}
=\mathcal{L}_\text{B}\mathcal{Q}=\mathcal{L}_\text{B},
\label{eqn:ProjLiouv}
\end{equation}
and that we can always accomplish
\begin{equation}
\mathcal{P}\mathcal{L}_\text{SB}\mathcal{P}=0,
\label{eqn:PLP}
\end{equation}
by redefining the Hamiltonians $H_\text{S}$ and $H_\text{SB}$.

\section{Projection Method}
Let us now start the derivation of the master equation for 
$\rho_\text{S}(t)$ defined by (\ref{eqn:PartialTrace}).
Consider the initial-value problem for the total system,
\begin{equation}
\frac{d}{dt}\rho(t)=\mathcal{L}\rho(t),\qquad
\rho(0)=\rho_0,
\label{eqn:A1}
\end{equation}
where the initial density operator $\rho_0$ is \textit{not} 
assumed to be factorized like (\ref{eqn:FactorizedInitialState}).
By projecting the Liouville equation (\ref{eqn:A1}) onto the two 
subspaces defined by $\mathcal{P}$ and $\mathcal{Q}$, and noting 
the properties (\ref{eqn:ProjLiouv}) and (\ref{eqn:PLP}), one gets
\begin{subequations}
\label{eqn:A3}
\begin{align}
\frac{\d}{\d t}\mathcal{P}\rho&=\mathcal{L}_\mathrm{S}
\mathcal{P}\rho
+\lambda\mathcal{P}\mathcal{L}_\mathrm{SB}\mathcal{Q}\rho,
\label{eqn:A3A}\\
\frac{\d}{\d t}\mathcal{Q}\rho&=\mathcal{L}'_0\mathcal{Q}\rho
+\lambda\mathcal{Q}\mathcal{L}_\mathrm{SB}\mathcal{P}\rho,
\label{eqn:A3B}
\end{align}
\end{subequations}
respectively, where
\begin{equation}\label{eqn:cLp0}
\mathcal{L}'_0
=\mathcal{L}_0+\lambda{\mathcal
Q}\mathcal{L}_\mathrm{SB}\mathcal{Q}.
\end{equation}
By formally integrating out the second equation and plugging the
result into the first, one gets the following equation for the 
$\mathcal{P}$-projected operator in the interaction
picture \cite{ref:HaakeGeneralizedMasterEq},
\begin{equation}
\frac{\d}{\d t}\e^{-\mathcal{L}_\mathrm{S}t}\mathcal{P}\rho(t)
=\lambda^2\rint_0^t\d t'\,\e^{-\mathcal{L}_\mathrm{S}t}
\mathcal{P}\mathcal{L}_\mathrm{SB}\e^{\mathcal{L}_0'(t-t')}
\mathcal{L}_\mathrm{SB}\mathcal{P}\rho(t')
+\lambda \e^{-\mathcal{L}_\mathrm{S}t}
\mathcal{P}\mathcal{L}_\mathrm{SB}\e^{\mathcal{L}'_0t}
\mathcal{Q}\rho_0.
\label{eqn:NonMarkovME}
\end{equation}
This is an \textit{exact} and \textit{non-Markovian} master 
equation since  the first term on the right-hand side contains a 
memory integral.
The last term, on the other hand, represents the contribution 
arising from a possible initial correlation between system S and 
reservoir B\@.
In fact, this term is absent when the initial state $\rho_0$ is 
factorized like (\ref{eqn:FactorizedInitialState}) and the 
reference state of the projector is accordingly chosen as 
$\Omega_\text{B}=\rho_\text{B}$.
We are interested in the fate of this inhomogeneous term in the 
following arguments, when there is an initial correlation between 
system S and reservoir B\@.

\section{Van Hove's Limit}
Let us make the Markovian approximation here, since the exact master 
equation (\ref{eqn:NonMarkovME}) is usually too complicated to 
handle for general systems and the Markovian approximation is 
valid for various practical situations.
In this article, we apply van Hove's limit \cite{ref:VanHove} for 
this purpose.
That is, we take the weak-coupling limit $\lambda\to0$ and the 
long-time limit $t\to\infty$ at the same time, keeping the scaled 
time $\tau=\lambda^2t$ finite.
This rescaling in time eliminates the memory, and the system 
exhibits a Markovian dynamics in the scaled time $\tau$.

Now consider the density operator
\begin{equation}\label{eq:rhoIlambda}
\rho_\mathrm{I}^{(\lambda)}(\tau) =
\e^{-\mathcal{L}_\mathrm{S}\tau/\lambda^2}\mathcal{P}
\rho(\tau/\lambda^2),
\end{equation}
that satisfies, for any nonvanishing $\lambda$,
\begin{multline}
\frac{\d}{\d\tau}\rho_\mathrm{I}^{(\lambda)}(\tau)
=\rint_0^{\tau/\lambda^2}\d t\,
\e^{-\mathcal{L}_\mathrm{S}\tau/\lambda^2}\mathcal{P}
\mathcal{L}_\mathrm{SB}\e^{\mathcal{L}'_0(\tau/\lambda^2-t)}
\mathcal{L}_\mathrm{SB}\mathcal{P}\e^{\mathcal{L}_\mathrm{S}t}
\rho_\mathrm{I}^{(\lambda)}(\lambda^2t)\\
{}+\frac{1}{\lambda}\e^{-\mathcal{L}_\mathrm{S}\tau/\lambda^2}
\mathcal{P}\mathcal{L}_\mathrm{SB}
\e^{\mathcal{L}'_0 \tau/\lambda^2}\mathcal{Q}\rho_0,
\label{eqn:A6}
\end{multline}
with the initial condition
\begin{equation}
\rho_\mathrm{I}^{(\lambda)}(0)=\mathcal{P}\rho_0.
\label{eqn:condiniz}
\end{equation}
By integrating (\ref{eqn:A6}), one gets
\begin{multline}
\rho_\mathrm{I}^{(\lambda)}(\tau)
=\mathcal{P}\rho_0
+\rint_0^\tau \d\tau'
\rint_0^{\tau'/\lambda^2}\d t\,
\e^{-\mathcal{L}_\mathrm{S}\tau'/\lambda^2}
\mathcal{P}\mathcal{L}_\mathrm{SB}
\e^{\mathcal{L}'_0(\tau'/\lambda^2-t)}\mathcal{L}_\mathrm{SB}
\mathcal{P}\e^{\mathcal{L}_\mathrm{S}t}
\rho_\mathrm{I}^{(\lambda)}(\lambda^2t)\\
{}+\frac{1}{\lambda}\rint_0^\tau \d\tau'\,
\e^{-\mathcal{L}_\mathrm{S}\tau'/\lambda^2}
\mathcal{P}\mathcal{L}_\mathrm{SB}
\e^{\mathcal{L}'_0\tau'/\lambda^2}\mathcal{Q}\rho_0,
\label{eqn:integriniz}
\end{multline}
which is arranged into the form \cite{ref:Factorize}
\begin{equation}
\rho_\mathrm{I}^{(\lambda)}(\tau)
=\mathcal{P}\rho_0 +\sum_{m,n}\rint_0^{\tau}\d\tau'\,
\e^{i(\omega_m-\omega_n)\tau'/\lambda^2}
\mathcal{K}_{mn}^{(\lambda)}(\tau-\tau')
\rho_\mathrm{I}^{(\lambda)}(\tau')
+\mathcal{I}^{(\lambda)}(\tau),
\label{eqn:ReducedDynamics}
\end{equation}
where
\begin{subequations}
\begin{align}
\mathcal{K}_{mn}^{(\lambda)}(\tau)
&=\mathcal{P}\tilde{Q}_m\mathcal{L}_\mathrm{SB}
\mathcal{R}_m^{(\lambda)}(\tau)\mathcal{L}_\mathrm{SB}
\tilde{Q}_n\mathcal{P},\label{eqn:Memory}\displaybreak[0]\\
\mathcal{I}^{(\lambda)}(\tau)
&=\lambda\sum_m\mathcal{P}\tilde{Q}_m\mathcal{L}_\mathrm{SB}
\mathcal{R}_m^{(\lambda)}(\tau)\mathcal{Q}\rho_0.
\label{eqn:CorrelationTerm}
\end{align}
\end{subequations}
The important feature is that the kernel 
$\mathcal{K}_{mn}^{(\lambda)}(\tau)$ and the contribution of the 
initial correlation $\mathcal{I}^{(\lambda)}(\tau)$ are both 
governed by the operator
\begin{equation}
\mathcal{R}_m^{(\lambda)}(\tau)
=\rint_0^{\tau/\lambda^2}\d t\,\mathcal{Q}
\e^{(\mathcal{L}_0'+i\omega_m)t}.
\label{eqn:KEY}
\end{equation}
Therefore, the van Hove limit of this key operator 
$\mathcal{R}_m^{(\lambda)}(\tau)$ completes the van Hove limit of 
the master equation (\ref{eqn:ReducedDynamics}).

\section{Main Theorem}
As we will see later, the van Hove limit of 
$\mathcal{R}_m^{(\lambda)}(\tau)$ results in
\begin{equation}
\mathcal{R}_m^{(\lambda)}(\tau)
\xrightarrow{\lambda\to0}
-\frac{\mathcal{Q}}{\mathcal{L}_0+i\omega_m-0^+}\quad(\tau>0),
\label{eqn:KeyFormula}
\end{equation}
and it immediately leads us to the conclusion that, in van Hove's 
limit, the memory kernel $\mathcal{K}_{mn}^{(\lambda)}(\tau)$ in 
(\ref{eqn:Memory}) is reduced to a Markovian generator
\begin{equation}
\mathcal{K}_{mn}^{(\lambda)}(\tau)
\xrightarrow{\lambda\to 0}
\mathcal{K}_{mn}^{(0)}
=-\mathcal{P}\tilde{Q}_m\mathcal{L}_\mathrm{SB}
\frac{\mathcal{Q}}{\mathcal{L}_0+i\omega_m-0^+}
\mathcal{L}_\mathrm{SB}\tilde{Q}_n\mathcal{P} ,
\end{equation}
while the correlation term $\mathcal{I}^{(\lambda)}(\tau)$ in
(\ref{eqn:CorrelationTerm}) disappears
\begin{eqnarray}
\mathcal{I}^{(\lambda)}(\tau)
\xrightarrow{\lambda\to 0}0,
\end{eqnarray}
so that the reduced dynamics (\ref{eqn:ReducedDynamics}) becomes
\begin{equation}
\rho_\mathrm{I}^{(\lambda)}(\tau)
\xrightarrow{\lambda\to0}\mathcal{P}\rho_0
+\sum_{m,n}\rint_0^{\tau}\d\tau'\,\delta_{mn}
\mathcal{K}_{mn}^{(0)} \rho_\mathrm{I}(\tau')
=\mathcal{P}\rho_0
+\rint_0^{\tau}\d\tau'\,\mathcal{K}\rho_\mathrm{I}(\tau').
\end{equation}
The master equation in van Hove's limit reads therefore 
\begin{subequations}
\label{eqn:ME}
\begin{equation}
\frac{\d}{\d\tau}\rho_\mathrm{I}(\tau)
=\mathcal{K}\rho_\mathrm{I}(\tau),\quad
\mathcal{K} =\sum_m
\mathcal{K}_{mm}^{(0)}
=-\sum_m\mathcal{P}\tilde{Q}_m\mathcal{L}_\mathrm{SB}
\frac{\mathcal{Q}}{\mathcal{L}_0+i\omega_m-0^+}
\mathcal{L}_\mathrm{SB}\tilde{Q}_m\mathcal{P},
\end{equation}
with the initial condition
\begin{equation}\label{eqn:diffeq}
\rho_\mathrm{I}(0) =\mathcal{P}\rho_0
=\tr_\mathrm{B}\{\rho_0\}\otimes\Omega_\mathrm{B}.
\end{equation}
\end{subequations}
That is, even if the initial state $\rho_0$ is not in a factorized
form, but rather there is entanglement, or simply a classical
correlation, between system S and reservoir B, all correlations
disappear in van Hove's limit and system S behaves as if the total
system started from the factorized initial state 
(\ref{eqn:diffeq}) with a reservoir state $\Omega_\mathrm{B}$
specified below.

The key formula (\ref{eqn:KeyFormula}), and therefore the master 
equation (\ref{eqn:ME}) in van Hove's limit, are proved under the 
following assumptions:
\begin{enumerate}
\item[(i)] the initial (correlated) state of the total system, 
$\rho_0$, belongs to a single sector.
That is, $\rho_0$ is given in the form
\begin{equation}
\rho_0=\sum_iL_i(1_\mathrm{S}\otimes\Omega_\mathrm{B})L_i^\dag,
\label{eqn:CondInitialState}
\end{equation}
where $1_\text{S}\otimes\Omega_\text{B}$ is the state which 
specifies the sector and $L_i$'s are bounded operators;
\item[(ii)] the state $\Omega_\text{B}$ is mixing, and therefore 
the reservoir Liouvillian $\mathcal{L}_\text{B}$ bears a simple 
eigenvalue $0$.
We assume in addition that the remaining part of the spectrum of 
$\mathcal{L}_\text{B}$ is absolutely continuous;
\item[(iii)] the projection operator $\mathcal{P}$ adopted in 
(\ref{eqn:KeyFormula}) and (\ref{eqn:ME}) must be defined in 
terms of $\Omega_\mathrm{B}$ that is mixing and
``contained'' in the initial state $\rho_0$ of the form 
(\ref{eqn:CondInitialState}).
\end{enumerate}

\section{Mixing}
\label{sec:mixing}
The state $\Omega_\text{B}$ is called mixing with respect to the 
reservoir dynamics $\e^{\mathcal{L}_\text{B}t}$ iff
\begin{equation}
\tr_\text{B}\{X(t)Y\Omega_\text{B}\}
=\tr_\text{B}\{X\e^{\mathcal{L}_\text{B}t}Y\Omega_\text{B}\}
\xrightarrow{t\to\infty}
\tr_\text{B}\{X\Omega_\text{B}\}
\tr_\text{B}\{Y\Omega_\text{B}\}
\label{eqn:MixingB}
\end{equation}
for any bounded superoperators $X$ and $Y$ of the reservoir B, where 
$X(t)=\e^{-\mathcal{L}_\text{B}t}X\e^{\mathcal{L}_\text{B}t}$ 
\cite{ref:ArnoldAvez}.
One of the typical mixing states is the thermal equilibrium state 
of bosons at finite temperature $\Omega_\text{B}\sim 
e^{-\beta H_\text{B}}$ \cite{ref:Factorize}, which often appears 
in the literature.
Among other interesting mixing states, there are nonequilibrium 
steady states, where system B consists of two reservoirs with 
different temperatures and steady current flows between them.
Such a situation is also within our scope.

Note here that a state $\rho_0$ that refers [in the sense of 
(\ref{eqn:CondInitialState})] to two (or more) different mixing 
states of an infinite reservoir cannot be a physical state, since 
it is the superposition of states belonging to different 
inequivalent sectors.
(Imagine, for example, the superposition of states with different 
temperatures.)
Hypothesis (i) is therefore reasonable from this point of view.

The important consequence of mixing is that it implies the 
existence of the point spectrum $0$ of $\mathcal{L}_\text{B}$ 
\cite{ref:ArnoldAvez}.
Let us consider the time evolution, driven by the reservoir free 
Liouvillian $\mathcal{L}_\text{B}$, of the expectation value of 
any operator $D=\sum_jA_j\otimes X_j$, starting from the initial 
state $\rho_0$ of the kind (\ref{eqn:CondInitialState}), where 
$A_j$'s are operators of system S and $X_j$'s bounded operators 
of reservoir B\@.
Then, the mixing property (\ref{eqn:MixingB}) of 
$\Omega_\mathrm{B}$ yields
\begin{multline}
\tr\{D\e^{\mathcal{L}_\mathrm{B}t}\rho_0\}
=\sum_i\sum_j\tr_\mathrm{B}[X_j\e^{\mathcal{L}_\mathrm{B}t}
\tr_\mathrm{S}\{
A_jL_i(1_\mathrm{S}\otimes\Omega_\mathrm{B})L_i^\dag
\}]\displaybreak[0]\\
\xrightarrow{t\to\infty}
\sum_i\sum_j\tr_\mathrm{B}\{X_j\Omega_\mathrm{B}\}\tr\{
A_jL_i(1_\mathrm{S}\otimes\Omega_\mathrm{B})L_i^\dag
\}
=\tr[D(\tr_\mathrm{B}\{\rho\}\otimes\Omega_\mathrm{B})],
\label{eqn:DiagonalProjectionB}
\end{multline}
and in this sense, we have
\begin{equation}
\e^{\mathcal{L}_\mathrm{B}t}\rho_0
\xrightarrow{t\to\infty}
\tr_\mathrm{B}\{\rho_0\}\otimes\Omega_\mathrm{B}
\label{eqn:BathRelaxation}
\end{equation}
for any state $\rho_0$ belonging to the sector of 
$1_\mathrm{S}\otimes\Omega_\mathrm{B}$.
This shows that $\mathcal{L}_\text{B}$ bears one and the only 
simple eigenvalue (in the relevant sector), which is $0$.
It is essential to note that, if $\mathcal{L}_\text{B}$ admits 
more simple eigenvalues, such as for an ergodic state 
$\Omega_\text{B}$ \cite{ref:ArnoldAvez}, it is not possible to 
derive a master equation as we will see later.
Recall further that mixing does not exclude a singular 
continuous spectrum, as far as it is transient, i.e., 
Riemann--Lebesgue's lemma holds \cite{ref:ArnoldAvez}.
Hypothesis (ii) therefore assumes stronger conditions than mixing.

Finally, Eq.~(\ref{eqn:BathRelaxation}) also shows that the 
eigenprojection $\Pi_0$ belonging to the simple eigenvalue $0$,
\begin{equation}
\mathcal{L}_\text{B}\Pi_0
=\Pi_0\mathcal{L}_\text{B}
=0,
\end{equation}
acts on the state $\rho$ in the relevant sector as
\begin{equation}
\Pi_0\rho=\tr_\mathrm{B}\{\rho\}\otimes\Omega_\mathrm{B}.
\label{eqn:Pi0A}
\end{equation}
The projector $\mathcal{P}$ suggested in Hypothesis (iii) is 
nothing but this eigenprojection $\Pi_0$: the proper choice of 
the projector is
\begin{equation}
\mathcal{P}=\Pi_0,
\label{eqn:ProperP}
\end{equation}
where the reference state $\Omega_\text{B}$ should be endowed 
with the mixing property and contained in the initial state as 
(\ref{eqn:CondInitialState}).
A wrong projector yields secular terms and prevents us from 
deriving a master equation, as we will see in the next section.

\section{The Key Formula}
Let us finally prove the key formula (\ref{eqn:KeyFormula}) with 
the projector (\ref{eqn:ProperP}).
It requires a few steps, since the spectral property of 
$\mathcal{L}_0'$ given in (\ref{eqn:cLp0}) is not known.
We start by noting that 
$\mathcal{Q}\e^{\mathcal{L}_0't}
=\mathcal{Q}\e^{\mathcal{L}t}\mathcal{Q}$ and hence 
$\|\mathcal{Q}\e^{\mathcal{L}_0't}\|
=\|\mathcal{Q}\e^{\mathcal{L}t}\mathcal{Q}\|
\le\|\mathcal{Q}\|^2\cdot\|\e^{\mathcal{L}t}\|=1$.
One therefore obtains
\begin{equation}
\lambda^n\mathcal{R}_m^{(\lambda)}(\tau)
=\lambda^n\rint_0^{\tau/\lambda^2}\d t\,\mathcal{Q}
\e^{(\mathcal{L}_0'+i\omega_m)t}
\to0\quad\mathrm{as}\quad\lambda\to0\quad(n>2)
\label{eqn:Highest}
\end{equation}
in van Hove's limit, irrespective of the spectrum of 
$\mathcal{L}_0'$.

Second, the convolution
\begin{equation}
\rint_0^t\d t'\,\e^{\mathcal{L}_0(t-t')}
\mathcal{Q}\mathcal{L}_\mathrm{SB}
\mathcal{Q}\e^{\mathcal{L}_0't'}
\label{eqn:Convolution}
\end{equation}
is bounded for any $t$, \textit{provided the projector is 
properly chosen as $\mathcal{P}=\Pi_0$ and the point spectrum of 
$\mathcal{L}_0$ is removed by the projection 
$\mathcal{Q}=1-\Pi_0$}.
See Ref.\ \cite{ref:Factorize}.

Now, by using
\begin{equation}
\e^{\mathcal{L}_0't} =\e^{\mathcal{L}_0t} +\lambda\rint_0^t\d
t'\,\e^{\mathcal{L}_0(t-t')}\mathcal{Q}
\mathcal{L}_\mathrm{SB}\mathcal{Q}\e^{\mathcal{L}_0't'},
\label{eqn:Perturbation}
\end{equation}
we expand the relevant quantity (\ref{eqn:KEY}) as
\begin{equation}
\mathcal{R}_m^{(\lambda)}(\tau) =\rint_0^{\tau/\lambda^2}\d
t\,\mathcal{Q} \e^{(\mathcal{L}_0+i\omega_m)t}
+\lambda\rint_0^{\tau/\lambda^2}\d t\rint_0^t\d t'\,
\e^{(\mathcal{L}_0+i\omega_m)t}\e^{-\mathcal{L}_0t'}\mathcal{Q}
\mathcal{L}_\mathrm{SB}\mathcal{Q}\e^{\mathcal{L}_0't'}.
\label{eqn:Expansion}
\end{equation}
Let us split the first term into two contributions, the discrete 
and the continuous spectra of $\mathcal{L}_0$, by the projection 
operators $\Pi_0$ and $1-\Pi_0$.
For $\tau>0$, one gets
\begin{align}
&\rint_0^{\tau/\lambda^2}\d t\,\mathcal{Q}
\e^{(\mathcal{L}_0+i\omega_m)t}\nonumber\\
&\quad=\rint_0^{\tau/\lambda^2}\d
t\,\mathcal{Q}\Pi_0\e^{(\mathcal{L}_\mathrm{S}+i\omega_m)t}
+\rint_0^{\tau/\lambda^2}\d t\,\mathcal{Q}(1-\Pi_0)
\e^{(\mathcal{L}_0+i\omega_m)t}\nonumber\\
&\quad=\frac{\tau}{\lambda^2}\mathcal{Q}\Pi_0\tilde{Q}_m 
+\sum_{n\neq m}\mathcal{Q}\Pi_0\tilde{Q}_n
\frac{\e^{i(\omega_m-\omega_n)\tau/\lambda^2}-1}%
{i(\omega_m-\omega_n)}
+\mathcal{Q}(1-\Pi_0)
\frac{\e^{(\mathcal{L}_0+i\omega_m)\tau/\lambda^2}-1}%
{\mathcal{L}_0+i\omega_m-0^+},
\label{eqn:ContPointSpec}
\end{align}
which shows that the only possible divergence of the relevant 
operator in (\ref{eqn:KEY}) in van Hove's limit $\lambda\to0$ 
\textit{comes from the point spectrum of $\mathcal{L}_0$} (i.e.\ %
the first term of the last expression).
\textit{If $\mathcal{P}=\Pi_0$, however, the divergent term 
disappears due to $\mathcal{Q}\Pi_0=0$: the contributions of the 
point spectrum are properly removed by $\mathcal{Q}=1-\Pi_0$.}
The correct projection operator $\mathcal{P}=\Pi_0$ allows us to 
arrange the second term in (\ref{eqn:Expansion}) into
\begin{multline}
\lambda\rint_0^{\tau/\lambda^2}\d t\rint_0^t\d t'\,
\e^{(\mathcal{L}_0+i\omega_m)t}\e^{-\mathcal{L}_0t'}\mathcal{Q}
\mathcal{L}_\mathrm{SB}\mathcal{Q}\e^{\mathcal{L}_0't'}
\displaybreak[0]\\
=\lambda\frac{\mathcal{Q}}{\mathcal{L}_0+i\omega_m-0^+}
\e^{i\omega_m\tau/\lambda^2} \rint_0^{\tau/\lambda^2}\d t\,
\e^{\mathcal{L}_0(\tau/\lambda^2-t)}\mathcal{Q}
\mathcal{L}_\mathrm{SB}\mathcal{Q}\e^{\mathcal{L}_0't}\\
{}-\lambda\frac{\mathcal{Q}}{\mathcal{L}_0+i\omega_m-0^+}
\mathcal{L}_\mathrm{SB}\rint_0^{\tau/\lambda^2}\d t\,\mathcal{Q}
\e^{(\mathcal{L}_0'+i\omega_m)t},
\label{eqn:HigherTerm}
\end{multline}
and we arrive at the recurrence formula
\begin{align}
\mathcal{R}_m^{(\lambda)}(\tau)
={}&\mathcal{Q}
\frac{\e^{(\mathcal{L}_0+i\omega_m)\tau/\lambda^2}-1}%
{\mathcal{L}_0+i\omega_m-0^+}\nonumber\displaybreak[0]\\
&{}+\lambda\frac{\mathcal{Q}}{\mathcal{L}_0+i\omega_m-0^+}
\e^{i\omega_m\tau/\lambda^2} \rint_0^{\tau/\lambda^2}\d t\,
\e^{\mathcal{L}_0(\tau/\lambda^2-t)}\mathcal{Q}
\mathcal{L}_\mathrm{SB}\mathcal{Q}\e^{\mathcal{L}_0't}\nonumber\\
&{}-\lambda\frac{\mathcal{Q}}{\mathcal{L}_0+i\omega_m-0^+}
\mathcal{L}_\mathrm{SB}\mathcal{R}_m^{(\lambda)}(\tau)\quad
(\tau>0).
\label{eqn:Recurrence}
\end{align}
The integral in the second term is the convolution 
(\ref{eqn:Convolution}), which is bounded for 
$\tau/\lambda^2\to\infty$ since the point spectrum of 
$\mathcal{L}_0$ is removed by the projection $\mathcal{Q}=1-\Pi_0$.
Therefore, by iterating the expansion (\ref{eqn:Recurrence}) 
twice and by noting
\begin{equation}
\lim_{t\to\pm\infty}
(1-\Pi_0)\frac{\e^{(\mathcal{L}_0+i\omega_m)t}}%
{\mathcal{L}_0+i\omega_m\mp0^+}=0
\end{equation}
together with (\ref{eqn:Highest}), we derive the key formula 
(\ref{eqn:KeyFormula}).

In summary, the correct projector $\mathcal{P}=\Pi_0$ helps us to 
manage the subtle point spectrum $0$ of $\mathcal{L}_\mathrm{B}$.
With a wrong projector, we would fail to appropriately remove it, 
Eq.\ (\ref{eqn:KeyFormula}) would not hold anymore, and 
$\mathcal{R}_m^{(\lambda)}(\tau)$ would diverge in van Hove's limit.
This is because the wrong operator projects the reservoir onto a 
wrong state, so that the system evolution is frustrated, and 
develops secular terms in $\tau/\lambda^2$ [like in Eq.\ %
(\ref{eqn:ContPointSpec})].

As stressed in Sec.\ 6, the reference state of the reservoir must 
be mixing in order that no discrete eigenvalue different from $0$ 
exists.
Otherwise, the point spectrum (except $0$) is out of control, 
since we do not know the eigenprojections other than $\Pi_0$, and 
again the emergence of  secular terms is inevitable.
Moreover, it should be noticed that the mixing property of the 
reservoir is crucial even for a \textit{factorized} initial state 
like (\ref{eqn:FactorizedInitialState}).

\section{Concluding Remarks}
In this article, we described the derivation of master equations 
via Nakajima--Zwanzig's projection operator method in the 
presence of the initial correlations with a reservoir.
In van Hove's limit, the initial correlation disappears and the 
total system behaves as if it started from a factorized initial 
condition.
It was demonstrated that (even for a factorized initial 
condition) the choice of the projection operator is crucial and a 
wrong projection provokes secular terms.
The mixing property of the reservoir is essential to the 
derivation of master equations.

The factorization of the initial state in van Hove's limit 
implies that the total system is \textit{factorized at any 
moment} in the scaled time $\tau$, since the ``initial time'' 
$t=0$ has no particular status.
It is actually possible to prove it, i.e.\ \cite{ref:Factorize}
\begin{equation}
\lim_{\lambda\to0}\mathcal{Q}\rho(\tau/\lambda^2)=0.
\end{equation}
It is worth noting that the interaction between system S and
reservoir B is not essential to the factorization: \textit{free 
evolution eliminates the correlation, and the reservoir relaxes 
into the mixing state $\Omega_\mathrm{B}$}. Indeed, for any state 
$\rho_0$ of the total system of the type 
(\ref{eqn:CondInitialState}) with a mixing state 
$\Omega_\text{B}$, we have
\begin{equation}
\e^{\mathcal{L}_0t}\rho_0
\xrightarrow{t\to\infty}
\e^{\mathcal{L}_\mathrm{S}t}\Pi_0\rho_0
=\e^{\mathcal{L}_\mathrm{S}t}
\tr_\mathrm{B}\{\rho_0\}\otimes\Omega_\mathrm{B},
\label{eqn:FreeFactorization}
\end{equation}
where the contribution of the continuous spectrum decays
out due to the mixing property. 
[See Eq.~(\ref{eqn:BathRelaxation}).]
In the rescaled time $\tau$, the factorization is very rapid 
(order $\lambda^2$ in the scaled time $\tau$), and the total
system looks factorized at any moment.
The timescales of the factorization and the mixing are related 
not only to the spread of the initial correlation and the 
perturbation from the mixing state but also to the size of the 
relevant reservoir observables.
These issues are discussed in detail in Ref.~\cite{ref:Factorize}.

\section*{Acknowledgements}
This work is in collaboration with S. Tasaki, P. Facchi, 
G. Kimura, H. Nakazato, I. Ohba, and S. Pascazio.
The author thanks D.~A.\ Lidar and A.\ Shabani for discussions.
This work is partly supported by the bilateral Italian--Japanese 
Projects II04C1AF4E on ``Quantum Information, Computation and 
Communication'' of the Italian Ministry of Instruction, 
University and Research, and 15C1 on ``Quantum Information and 
Computation'' of the Italian Ministry for Foreign Affairs, by the 
Grant for The 21st Century COE Program at Waseda University, the 
Grant-in-Aid for the COE Research at Waseda University 
(No.~13CE2003), and the Grant-in-Aids for Scientific Research on 
Priority Areas (No.~14077219) and (No.~13135221), from MEXT, 
Japan, and by Grant-in-Aids for Scientific Research (C) 
(No.~14540280 and No.\ 17540365) from JSPS\@.

\end{document}